\documentclass[aps,reprint,amsmath,amssymb]{revtex4-2}
\usepackage{dcolumn}
\usepackage{amsthm}
\usepackage{graphicx,subfigure}
\usepackage{epsfig}
\usepackage{float}
\usepackage{scalerel}
\usepackage{dblfloatfix}
\usepackage{caption}
\usepackage[utf8]{inputenc}
\usepackage{xcolor, soul}
\sethlcolor{yellow}

\begin{document}

\preprint{APS/123-QED}
\title{Kosambi-Cartan-Chern (KCC) Perspective on Chaos: Unveiling Hidden Attractors in Nonlinear Autonomous Systems.}

\author{Somnath Roy}
\email{roysomnath63@gmail.com}
\affiliation{ Department of Physics,Jadavpur university, Kolkata-700075}
\author{Anirban Ray}%
\email{anirban.chaos@gmail.com}
\affiliation{ Department of Physics,Gour Mahavidyalaya,Mangalbari, Malda-732142}
\author{A Roy Chowdhury}%
\email{arc.roy@gmail.com (corresponding author)}
\affiliation{Department of Physics,Jadavpur university,Kolkata-700075 }

\begin{abstract}
\noindent This article confronts the formidable task of exploring chaos within hidden attractors in nonlinear 3-D autonomous systems, highlighting the lack of established analytical and numerical methodologies for such investigations. As the basin of attraction does not touch the unstable manifold ,there are no straightforward numerical processes to detect those attractors and one has to implement special numerical-analytical strategy.In this article we present an alternative approach that allows us to predict the basin of attraction associated with hidden attractors, overcoming the existing limitations. The method discussed here based on KCC theory (Kosambi-Cartan-Chern) which enable us to conduct a comprehensive theoretical analysis by means of evaluating geometric invariants and instability exponents, thereby delineating the regions encompassing chaotic and periodic zones. Our analytical predictions are thoroughly validated by numerical results.
\\\\


\end{abstract}
\maketitle

\section*{I:INTRODUCTION}
From the latter half of the century onwards, the enigmatic world of chaotic oscillators has beckoned researchers, casting a spell of fascination and intrigue that continues to deepen.In that period chaotic self-excited attractors were found numerically \cite{ueda,lorenz} by standard numerical procedure ,in which the trajectory starting from a unstable manifold in the vicinity of equilibrium reaches a state of oscillation after an intermediate transient process, thereby localization of basin of attraction of these kind of self excited chaotic oscillators is somewhat systematic. Since that time, an abundance of research has been diligently carried out in this fascinating realm \cite{feudel1,feudel2,feudel3,feudel4,kapitaniak1,kapitaniak2,kapitaniak3,kapitaniak4,kapitaniak5,kapitaniak6,kapitaniak7}.In recent times, a significant surge of scholarly attention has been directed towards a remarkable development, spurred by the seminal work of Kuznetsov and Lenov \cite{Kuznetsov1}. This notable advancement unveils a distinctive class of attractors, characterized as \textit{hidden attractors} \cite{Kuznetsov2,Kuznetsov3,Kuznetsov4,Kuznetsov5,kapitaniak8,Kuznetsov6}, where the basin of attraction remains devoid of any intersection with the unstable manifold of equilibrium points.For instance, within the framework of dynamical systems, cases characterized by the absence of any equilibrium points or the existence of only a single equilibrium point engender the emergence of chaotic hidden oscillations \cite{nose,hoover,wei,jafari,arc,wang,kingini}. In such intricate scenarios, the computational and analytical endeavour associated with the localization of the basin of attraction assumes a distinctly formidable character, primarily due to the inherent challenge posed by the unavailability of a priori equilibrium point information.
Extensive research efforts in recent years have been directed towards the localization of hidden attractors, employing both numerical and analytical approaches. Numerical procedures encompass methodologies such as homotopy and continuation techniques\cite{Kuznetsov7}, while analytical methods are rooted in the construction of Lyapunov functions and an assessment of the dissipative properties inherent to the system\cite{kapitaniak9}. Additionally, a novel approach has emerged involving the construction of perpetual points\cite{kapitaniak10}.\\

However, despite these significant advancements, the field currently lacks an analytical method capable of predicting the basin of attraction for hidden oscillators. Addressing this critical gap in the existing body of knowledge constitutes the central focus of our article. In our pursuit, we employ a novel geometrodynamical approach, rooted in the Kosambi-Cartan-Chern (KCC) theory \cite{kosambi,cartan,chern}, which is founded on the principles of Finsler space theory \cite{Antonelli,Rund}. This approach holds the promise of shedding new light on the intricate problem of predicting basin of attraction for hidden oscillators.\\

While global stability analysis of chaotic systems can traditionally be characterized through Lyapunov stability assessments, it is worth noting that the numerical calculation of Lyapunov exponents, often necessitating substantial computational resources, can pose considerable challenges in practice.\\

Alternatively, within the framework of the Kosambi-Cartan-Chern (KCC) theory, one finds a compelling avenue for stability analysis. Among the five geometric invariants offered by the KCC theory, the second invariant, namely the \textit{deviation curvature tensor} ($P_{j}^i$), provides a valuable insight into the bunching and dispersing behavior of nearby trajectories, near the arbirarily choosen initial positions. This approach offers an alternative perspective on stability analysis, commonly known as Jacobian stability. Jacobi stability analysis, as elucidated in detail by Harko et al.\cite{harko1}, has become a pivotal tool in the analysis of complex dynamical systems. Its applicability extends across a diverse spectrum of systems, including but not limited to the renowned Lorenz, Rossler,Rabinovich–Fabrikant and Chen chaotic systems \cite{harko2,huang,gupta1,gupta2,gupta3}. Furthermore, Jacobi stability analysis has found successful application in the study of biological systems, encompassing areas such as cell dynamics \cite{Antonelli2,gupta4}, prey-predator models\cite{munteanu}, and competitive models\cite{Yamasaki}. Its versatility is further underscored by its effectiveness in the investigation of various bifurcation phenomena\cite{Yongjian,Yamasaki2}, establishing it as an indispensable analytical framework in the realm of nonlinear dynamics.\\

In this article, our exploration focuses on two distinct autonomous 3-D systems. The first system, characterized by a sole stable fixed point\cite{wang}, presents an intriguing dynamic. On the other hand, the second system,with out any fixed points\cite{jafari2}, represents a rare and challenging dynamical scenario. Remarkably, both systems exhibit hidden chaotic oscillations.Our primary objective revolves around the analytical depiction of the basin of attraction for these oscillatory systems. We achieve this by meticulously evaluating deviation vectors originating from arbitrary initial points within the phase space, offering valuable insights into the hidden dynamics at play.

\renewcommand{\theequation}{2.\arabic{equation}}
\setcounter{equation}{0}
\section*{II: REVIEW OF KCC-THEORY AND JACOBI STABILITY}
We revoke the fundamentals of KCC theory which will be used in the following sections. We follow the work of C.G B\"{o}hmer et al \cite{harko1}. Let p=$(x,y) \in T\mathcal{M}$ where $x=(x^1,x^2,....,x^n)$, $y=(y^1,y^2,....,y^n)$ and $T\mathcal{M}$ be the tangent bundle of the smooth n-dimensional manifold $\mathcal{M}=\mathbb{R}^n$. Now consider a open connected subset  $\Omega \in \mathbb{R}^n\times\mathbb{R}^n\times\mathbb{R}$ and $(x,y,t) \in \Omega$ where we consider a system second-order differential equations in the form of

\begin{equation}
 \frac{d^2x^i}{dt^2}+2G^i(x,y,t)=0  ~~~~~~i \in \{1,2,....,n\}
 \label{eq2.1}
\end{equation}

where $G^i$ is a smooth function of local coordinates defined on $T\mathcal{M}$.Now by defining time independent coordinate transformation $\tilde{x}^i=\tilde{x}^i(x^1,x^2,...,x^n)$ and $\tilde{y}^i=\frac{\partial\tilde{x}^i}{\partial{x^j}}y^j$, the equivalent vector field $V$ on $T\mathcal{M}$ of Eq.(\ref{eq2.1}) is given by

\begin{equation}
 V=y^i\frac{\partial}{\partial x^i}-2G^i(x^j,y^j,t)\frac{\partial}{\partial y^i}
 \label{eq2.2}
\end{equation}
from which one can establish the nonlinear connection $N_{i}^j$ defined by \cite{harko1}

\begin{equation}
 N_{i}^j=\frac{\partial G^i}{\partial y^j}.
 \label{eq2.3}
\end{equation}

We can proceed further to obtain covariant differential of vector field $\xi^i\subseteq \Omega$ as \cite{harko1}
\begin{equation}
 \frac{D\xi^i}{dt}=\frac{d\xi^i}{dt}+N_{j}^i\xi^j.
 \label{eq2.4}
\end{equation}

Substituting $\xi^i=y^i$ we can generation

\begin{equation}
 \frac{D y^i}{dt}=N_{j}^i y^j-2G^i=-\epsilon^i,
 \label{eq2.5}
\end{equation}
where $\epsilon^i$ is the contravariant vector field defined on$\Omega$ and known as \textit{first KCC invariant} which physically implicates the external force present in the system.\\

If we now deviate the trajectory of system Eq.(\ref{eq2.1}) from nearby ones according to

\begin{equation}
 \tilde x^i(t)=x^i(t)+\eta\xi^i(t)
 \label{eq2.6}
\end{equation}
with $\parallel\eta\parallel$ is treated as small perturbation parameter and $\xi^i$ as a component of contravariant vector field along the path $x^i$ and substituting the above equation into Eq.(\ref{eq2.1}) and taking the limit $\eta\longrightarrow0$, we can arrive to the variational equations \cite{harko1}

\begin{equation}
 \frac{d^2\xi^i}{dt^2}+2N_{j}^i\frac{d\xi^j}{dt}+2\frac{\partial G^i}{\partial x^j}\xi^j=0
 \label{eq2.7}
\end{equation}
with KCC covariant differential the above equation takes the form
\begin{equation}
 \frac{D^2\xi^i}{dt^2}=P_{j}^i\xi^j
 \label{eq2.8}
\end{equation}
where $P_{j}^i$ is defined as
\begin{equation}
 P_{j}^i=-2\frac{\partial G^i}{\partial x^j}-2G^l G_{jl}^i+y^l\frac{\partial N_{j}^i}{\partial x^l}+N_{l}^i N_{j}^l
 \label{eq2.9}
\end{equation}
The above equation Eq.(\ref{eq2.8}) is called \textit{Jacobi equation} of the second-order differential equation, and $P_{j}^i$ is symbolized as the \textit{second KCC invariant or the deviation curvature tensor}, with the Berwald connection denoted as $G_{jl}^i\equiv\frac{\partial N_{j}^i}{\partial y^l}$. There are three more KCC invariants \cite{harko1} which are excluded here due to the  motive our article.\\

Now to analyze Jacobi stability of the trajectories $x^i=x^i(t)$ of Eq.(\ref{eq2.1}) in the vicinity of a point $x^i(t_{0}=0)$ in the Euclidean space $(\mathbb{R}^n,\langle.,.\rangle)$, we have to study the behavior of the deviation vector $\xi^i$ which satisfies the initial conditions $\xi(0)=O$ and $\dot{\xi}(0)=W \neq O$, where $O$ be the null vector in $\mathbb{R}^n$. For arbitrary two vectors $\langle\langle X,Y\rangle\rangle \in \mathbb{R}^n$ we consider an adapted inner product $\langle\langle.,.\rangle\rangle$ of $\xi$ such that $\langle\langle X,Y\rangle\rangle:= \frac{1}{\langle W,W\rangle}\cdot\langle X,Y\rangle$.\\

\begin{figure}[h]
\begin{center}
\includegraphics[width=0.5\textwidth]{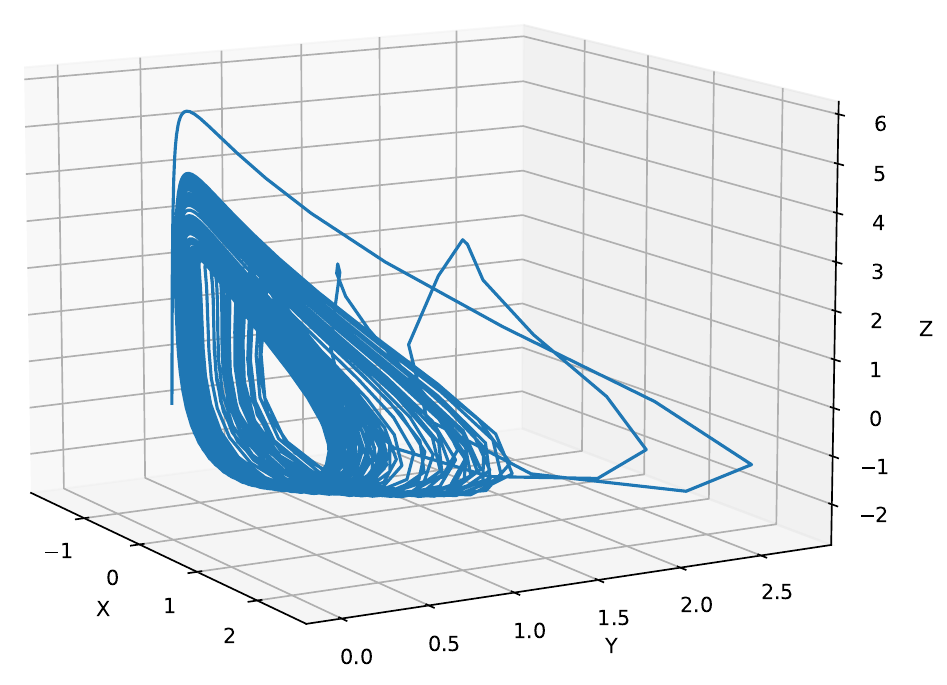}
\caption{Attractor of the system Eq.(\ref{eq3.1}-\ref{eq3.3}) with initial condition $(0,0,0)$}
\label{fig1}
\end{center}
\end{figure}

This enable us to describe the bunching and dispersing tendency of $\xi$ around $t_0=0$ as follows:  as $t\rightarrow 0^+$ if $\| \xi \| < t^2$, then the trajectories are bunching together, and if $\| \xi \| > t^2$ as $t\rightarrow 0^+$, the trajectories are dispersing.\\\\

\noindent\textbf{\ul{Definition:}} If Eq.(\ref{eq2.1}) satisfies the initial conditions mentioned above then the trajectories $x^i(t)$ are called \textit{Jacobi stable} if and only if the real part of the eigenvalues of the deviation curvature tensor $P_{j}^i(0)$ is negative.\\

In the upcoming sections, we undertake the analysis of two distinct systems, each exemplifying a unique dynamical characteristic—one with a single equilibrium point and the other devoid of any fixed points. Our objective is to demonstrate that the Kosambi-Cartan-Chern (KCC) analysis is adept at extracting the basin of attraction for both systems through an analytical framework. This framework is expressed in terms of a set of coupled differential equations governing the deviation vectors, denoted as $\xi^i$. By determining the solution of $\xi^i$, we subsequently define the instability exponent $\delta_i$, akin to the Lyapunov exponent, as a quantitative measure of the chaotic behavior. The deviation vector can be evaluated from its components by
\begin{equation}
\xi(t)=\frac{\sqrt{[\xi^1(t)]^2+[\xi^2(t)]^2}}{\sqrt{[\dot{\xi^1}(0)]^2+[\dot{\xi^2}(0)]^2}}.
\label{eq2.10}
\end{equation}
now the instability exponents are defined as

\begin{equation}
\delta_i(x^j,y^j,t)=\lim_{t \to \infty} \frac{1}{t}ln\bigg[\frac{\xi^i}{\xi_{i0}}\bigg]~~~~~~i=0,1
\label{eq2.11}
\end{equation}
and
\begin{equation}
\delta(x^j,y^j,t)=\lim_{t \to \infty} \frac{1}{t}ln\bigg[\frac{\xi}{\xi_{10}}\bigg],
\label{eq2.12}
\end{equation}

where $\dot{\xi^1}(0)=\xi_{10}$ and $\dot{\xi^2}(0)=\xi_{20}$. It is important to note that the instability exponent, in general, is a function of initial conditions $(x^j,y^j,t)$ and holds the potential to predict the basin. The notable advantage lies in having an analytical description of the instability exponent, in contrast to the Lyapunov exponent, which, in general, defies analytical calculation. In most cases, numerical techniques become imperative for the computation of Lyapunov exponents.
\begin{figure}[h]
\begin{center}
\includegraphics[width=0.4\textwidth]{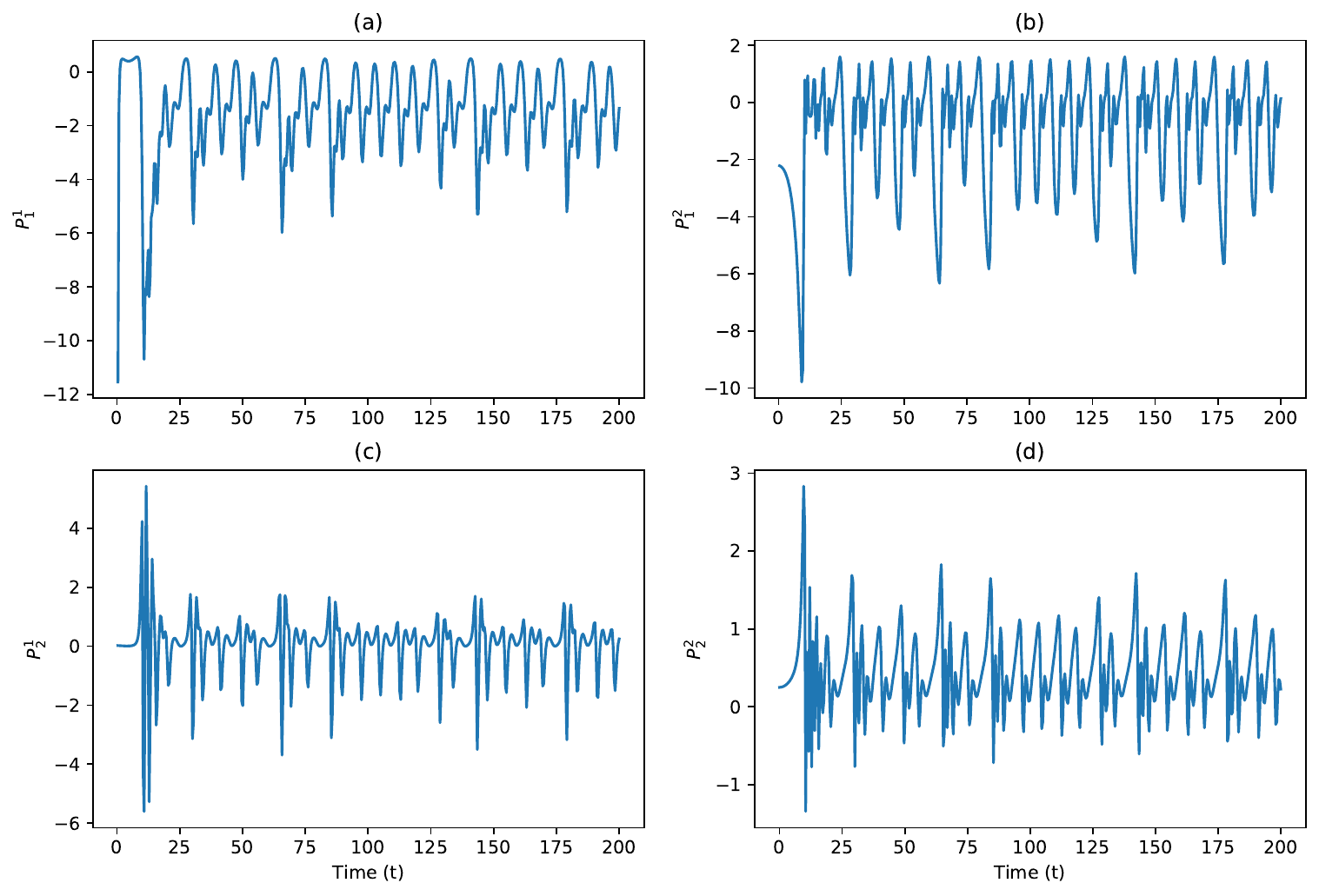}
\caption{Components of deviation curvature tensor $P_j^i$ Eq.(\ref{eq3.10}) as varies with time for the system Eq.(\ref{eq3.1}-\ref{eq3.3}) with initial condition (0,0,0). (a)$P_1^1$(t) vs time(t). (b)$P_2^1$(t) vs time(t). (c)$P_1^2$(t) vs time(t). (d)$P_2^2$(t) vs time(t) }
\label{fig2}
\end{center}
\end{figure}

\renewcommand{\theequation}{3.\arabic{equation}}
\setcounter{equation}{0}
\section*{IV: STABILITY OF A SYSTEM WITH SINGLE FIXED POINT}

\begin{eqnarray}
\dot{x}&=yz+a \label{eq3.1}\\
\dot{y}&=x^2-y \label{eq3.2}\\
\dot{z}&=1-4x \label{eq3.3}
\end{eqnarray}

This system has a single fixed point at $S(x_0,y_0,z_0)=(\frac{1}{4},\frac{1}{16},-16a)$.The attractor is shown in Fig.(\ref{fig1}) .Now defining $x=X^1,y=X^2,z=X^3; \dot{x}=Y^1,\dot{y}=Y^2,\dot{z}=Y^3$.By rearranging the equations, following the methodology detailed in the previous section, we can express the system as:

\begin{eqnarray}
&\frac{d^2X^1}{dt^2}+2G^1=0 \label{eq3.4}\\
&\frac{d^2X^2}{dt^2}+2G^2=0 \label{eq3.5}
\end{eqnarray}

where,
\begin{align}
G^1=&\frac{1}{2}\big[X^2(4X^1-1)+\frac{Y^2}{X^2}(a-Y^1)\big]\label{eq3.6}\\
G^2=&\frac{1}{2}\big[Y^2-2X^1Y^1\big]\label{eq3.7}
\end{align}

The components of the nonlinear connections are determined by,

\begin{equation}
\begin{split}
\begin{cases}
N_{1}^1&=-\frac{1}{2}\frac{Y^2}{X^2},~~~N_{1}^2=-\frac{1}{2}\frac{a-Y^1}{X^2},\\
N_{2}^1&=-X^1,~~~~N_{2}^2=\frac{1}{2}.
\label{eq3.8}
\end{cases}
\end{split}
\end{equation}

The components of berwald components are,

\begin{equation}
\begin{split}
\begin{cases}
G_{11}^1&=0,G_{12}^1=-\frac{1}{2X^2}, G_{21}^1=-\frac{1}{2X^2}, G_{22}^1=0,\\
G_{11}^2&=G_{12}^2=G_{21}^2=G_{22}^2=0.
\label{eq3.9}
\end{cases}
\end{split}
\end{equation}

Now by Eq.(\ref{eq2.9}) the components of deviation curvature tensor expressed as

\begin{equation}
\begin{cases}
\begin{split}
P_{1}^1&=-4X^2+\frac{1}{2X^2}\big[Y^2-2X^1Y^1\big]+\frac{1}{2}\bigg(\frac{Y^2}{X^2}\bigg)^2\\
+&\frac{1}{4}\bigg(\frac{Y^2}{X^2}\bigg)^2-\frac{X^1}{2}\big[\frac{a-Y^1}{X^2}\big],\\
P_{2}^1&=\big[\frac{Y^2}{(X^2)^2}(a-Y^1)-(4X^1-1)\big]+\frac{1}{2X^2}\big[X^2(4X^1-1)\\
+&\frac{Y^2}{X^2}(a-Y^1)\big]+\frac{Y^2}{2}\big[\frac{Y^1-a}{(X^2)^2}\big]-\frac{Y^2}{4(X^2)^2}\big[a-Y^1\big]\\
+&\frac{1}{4}\big[\frac{a-Y^1}{X^2}\big],\\
P_{1}^2&=Y^1+\frac{X^1Y^2}{2X^2}-\frac{X^1}{2},\\
P_{2}^2&=\frac{1}{4}-\frac{X^1}{2X^2}\big(a-Y^1\big).
\end{split}
\end{cases}
\label{eq3.10}
\end{equation}

\renewcommand\thesubfigure{(\alph{subfigure})}
\setcounter{subfigure}{0}
\begin{figure}[ht]
\centering
\subfigure[]{\includegraphics[width=0.25\textwidth]{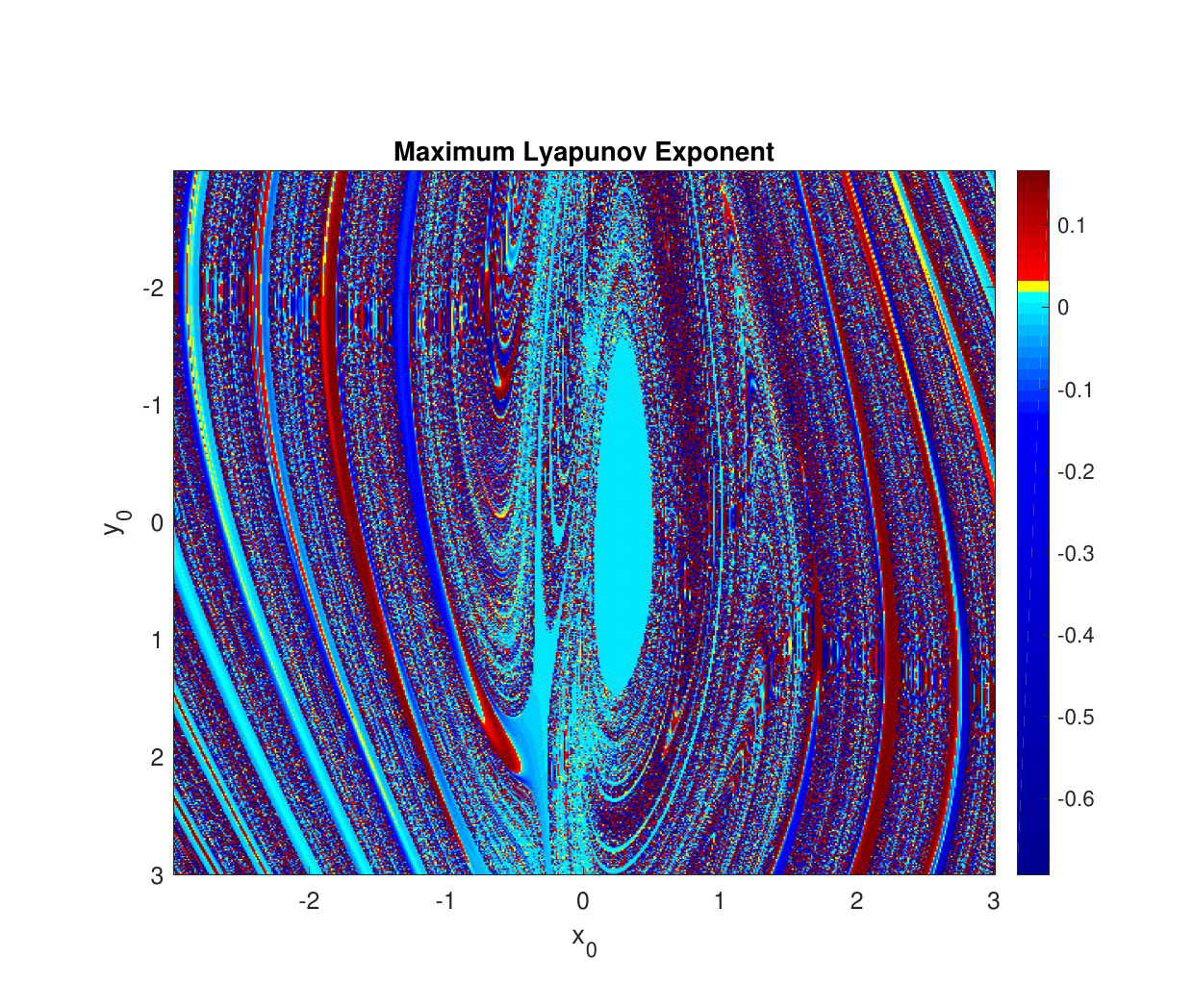}}\quad
\subfigure[]{\includegraphics[width=0.25\textwidth]{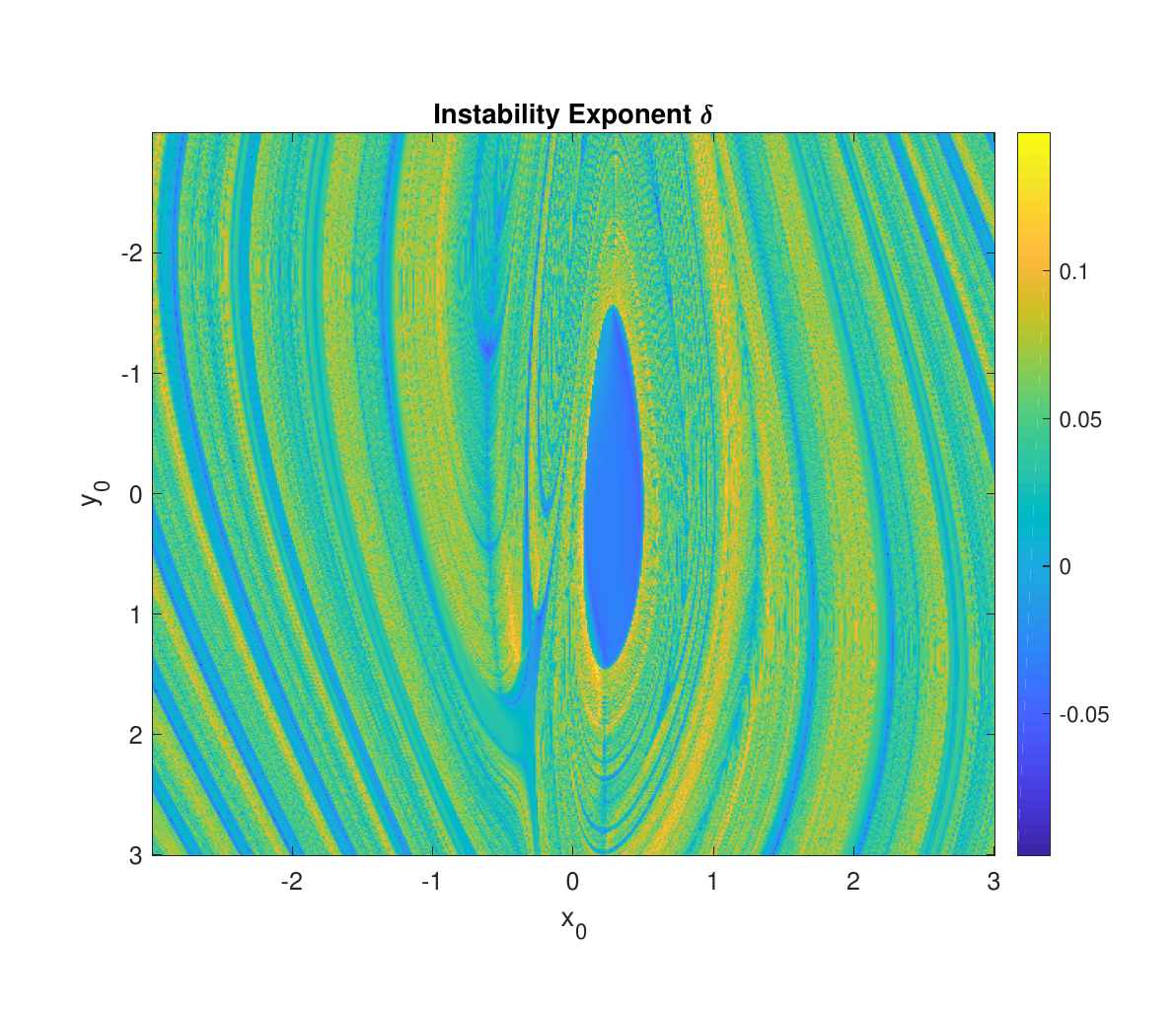}}
\caption{Basin of attraction of the system Eq.(\ref{eq3.1}-\ref{eq3.3}) (a)Basin ($x_0,y_0$) is numerically determined by the estimation of maximum Lyapunov exponent (b) Basin by calculating the instability exponent $\delta$ with help of deviation vector computed from Eq.(\ref{eq3.15}-\ref{eq3.16}). }
\label{fig3}
\end{figure}

Now we have all components of $P_j^i$ matrix which can be evaluated at fixed point $S(x_0,y_0,z_0)$

$\begin{bmatrix}
P_{1}^1 & P_{1}^2\\
P_{2}^1 & P_{2}^2
\end{bmatrix}_S$=$\begin{bmatrix}
-\frac{1}{4}-2a & 4a\\
-\frac{1}{8} & \frac{1}{4}-2a
\end{bmatrix}$,
gives the characteristics equation
\begin{equation}
\lambda^2+4a\lambda+4a^2-\frac{1}{16}+\frac{a}{2}=0.
\label{eq3.11}
\end{equation}
The eigenvalues of the above equation are
\begin{equation}
\lambda_{1,2}=-2a\pm\frac{1}{2}\sqrt{\frac{1}{4}-2a}.
\label{eq3.12}
\end{equation}
Clearly the equilibrium $S(x_0,y_0,z_0)$ is Jacobi unstable and its a saddle focus \cite{harko1}.
The deviation vectors $\xi^i$ can be written by Eq.(\ref{eq2.7}) as

\begin{align}
\frac{d^2\xi^1}{dt^2}-&\frac{Y^2}{X^2}\frac{d\xi^1}{dt}+(\frac{a-Y^1}{X^2})\frac{d\xi^2}{dt}+4X^2\xi^1\nonumber\\
&\bigg[(4X^1-1)-\frac{Y^2}{(X^2)^2}(a-Y^1)\bigg]\xi^2=0\label{eq3.13}
\end{align}
and
\begin{align}
\frac{d\xi^2}{dt}-2X^1\frac{d\xi^1}{dt}+\frac{d\xi^2}{dt}-2Y^1\xi^1=0\label{eq3.14}
\end{align}

Substituting the systems variables in the above equations take the form respectively as

\begin{align}
\frac{d^2\xi^1}{dt^2}-&\frac{x^2-y}{y}\frac{d\xi^1}{dt}-z\frac{d\xi^2}{dt}+4y\xi^1\nonumber\\
&\bigg[(4x-1)+\frac{z(x^2-y)}{y}\bigg]\xi^2=0\label{eq3.15}
\end{align}

and
\begin{align}
\frac{d\xi^2}{dt}-2x\frac{d\xi^1}{dt}+\frac{d\xi^2}{dt}-2(yz+a)\xi^1=0\label{eq3.16}
\end{align}

Equations (\ref{eq3.15}) and (\ref{eq3.16}) represent a system of coupled differential equations for $\xi^1$ and $\xi^2$, offering a depiction of the basin for the system described by Equations (\ref{eq3.1}-\ref{eq3.3}). The solutions to these equations can be obtained through numerical methods such as RK-4 or Euler. Subsequently, utilizing Equations (\ref{eq2.10}) and (\ref{eq2.12}), the instability exponent $\delta$ can be determined as a function of the system's variables. This approach offers a computationally more tractable alternative to the laborious task of numerically calculating Lyapunov exponents.In Fig.(\ref{fig3}), two basins are presented. One is computed by tracing the maximum Lyapunov exponent, while the other is determined by calculating the instability exponent using Eq.(\ref{eq2.12}). Remarkably, the figures exhibit a close alignment, providing an accurate visualization of the periodic and chaotic zones associated with the hidden attractor.

\subsection{Behaviour of the deviation vectors near the fixed point}
The linear stability analysis reveals the existence of a stable fixed point at $S$ within the system. However, when subjected to Jacobi stability analysis, the equilibrium point emerges as an unstable saddle focus. This intriguing contradiction motivates us to scrutinize the curvature variation around this point.At first Eq.(\ref{eq3.15} and \ref{eq3.16}) take the form of a linearly coupled equation near the fixed point

\begin{equation}
\frac{d^2\xi^1}{dt^2}+16a\frac{d\xi^2}{dt}+\frac{1}{4}\xi^1=0 \label{eq3.17}
\end{equation}
and
\begin{equation}
\frac{d\xi^2}{dt}-\frac{1}{2}\frac{d\xi^1}{dt}+\frac{d\xi^2}{dt}=0 \label{eq3.18}
\end{equation}

The deviation vectors are shown in Fig.() for $a=0.01$. From the information of $\xi^1$ and $\xi^2$ one can derive the curvature \cite{harko1} defined as

\begin{equation}
\kappa(S)=\frac{\dot{\xi^1}(t)\ddot{\xi^2}(t)-\dot{\xi^2}(t)\ddot{\xi^1}(t)}{\big([\xi^1(t)]^2+[\xi^2(t)]^2\big)^\frac{3}{2}}\label{eq3.19}
\end{equation}

\begin{figure}[ht]
\centering
\subfigure[]{\includegraphics[width=0.25\textwidth,height=0.25\textwidth]{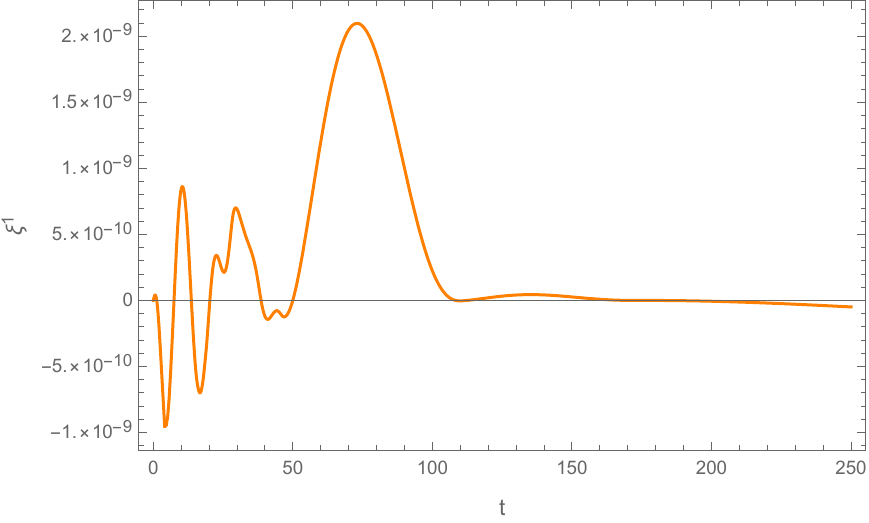}}
\subfigure[]{\includegraphics[width=0.25\textwidth,height=0.25\textwidth]{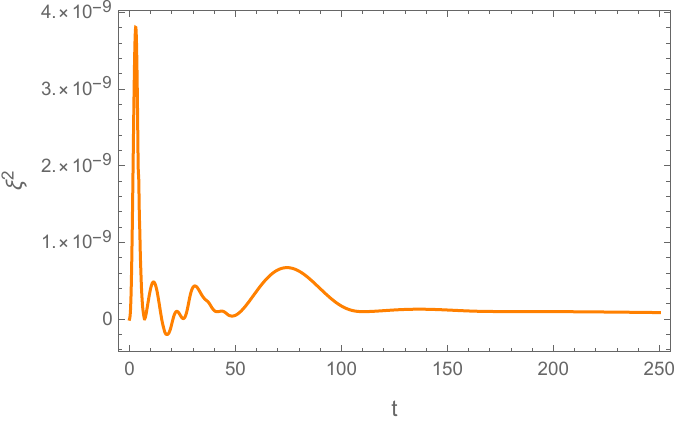}}
\subfigure[]{\includegraphics[width=0.25\textwidth,height=0.25\textwidth]{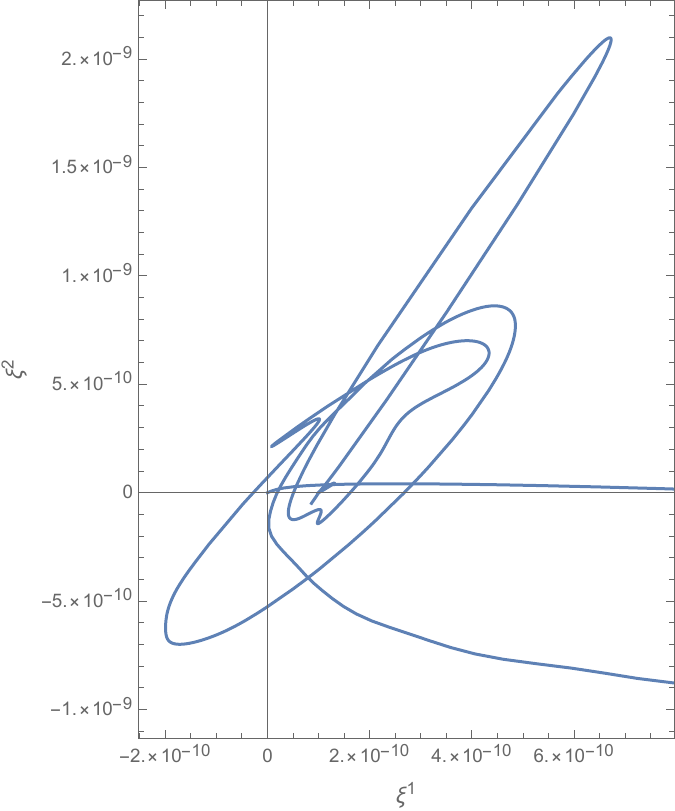}}
\caption{Deviations vectors are numerically calculated from Eq.(\ref{eq3.15}-\ref{eq3.16}) (a)Variation of deviation vector $\xi^1$ with time. Initial conditions are $\dot{\xi^1}(0)=10^{-10}$ and $\dot{\xi^2}(0)=10^{-10}$ (b) Variation of deviation vector $\xi^2$ with time (c)Phase portrait in $\xi^1$-$\xi^2$ plane}
\label{fig4}
\end{figure}

\begin{figure}[ht]
\centering
\subfigure[]{\includegraphics[width=0.25\textwidth]{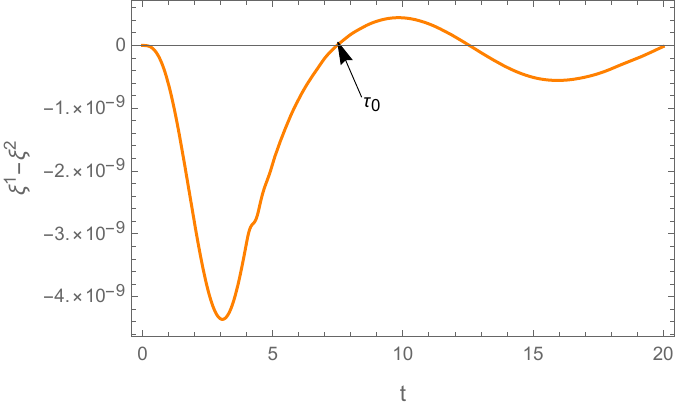}}
\hspace{2cm}
\subfigure[]{\includegraphics[width=0.25\textwidth]{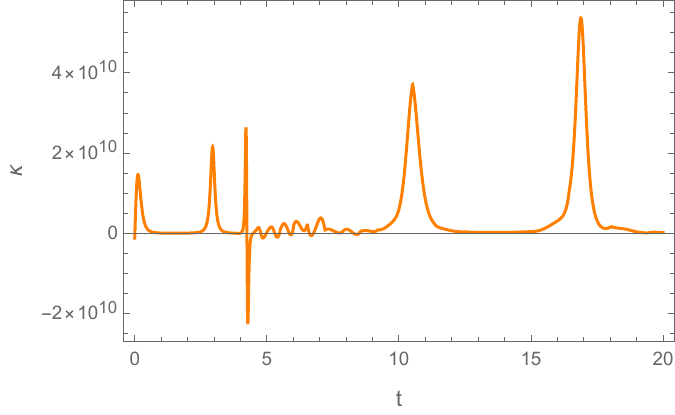}}
\caption{(a) Critical time $\tau_0$ at which the curvature $\kappa$ is zero, i.e $\xi^1(\tau_0)=\xi^2(\tau_0)$, is approximately 7.5. (b) Variation of curvature $\kappa$ with time (t). Clearly shows that the curvature changes its sign before the critical time, which is a quantitative indication of chaos. }
\label{fig5}
\end{figure}

The variation of $\xi^1$ and $\xi^2$ is depicted in Fig.(\ref{fig4}). Moving beyond the knowledge of deviation vectors, we proceed to illustrate the curvature variation using Eq.(\ref{eq3.19}) in Fig.(\ref{fig5}). It is widely recognized that the onset of chaos within the system can be identified by observing whether the curvature changes sign before a critical minimum time. This critical time, denoted as $\tau_0$, is determined by setting $\xi^1(t_0)=\xi^2(t_0)$, implying the curvature $\kappa(t_0)$ is zero. In Fig.(\ref{fig5}(a)), the numerical calculation yields a value of approximately 7.5 for $\tau_0$. Furthermore, in Fig.(\ref{fig5}(b)), it is evident that the curvature changes sign before $\tau_0$, indicating that the underlying evolution of this hidden attractor will exhibit chaotic behavior in the long term.

\renewcommand\thesubfigure{(\alph{subfigure})}
\setcounter{subfigure}{0}

\renewcommand\thesubfigure{(\alph{subfigure})}
\setcounter{subfigure}{0}

\begin{figure}[ht]
\centering
\subfigure[]{\includegraphics[width=0.25\textwidth]{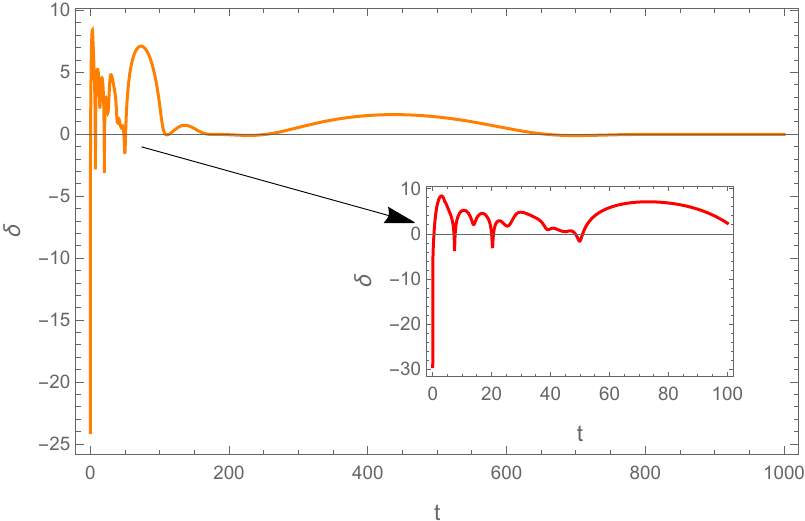}}
\hspace{2cm}
\subfigure[]{\includegraphics[width=0.25\textwidth]{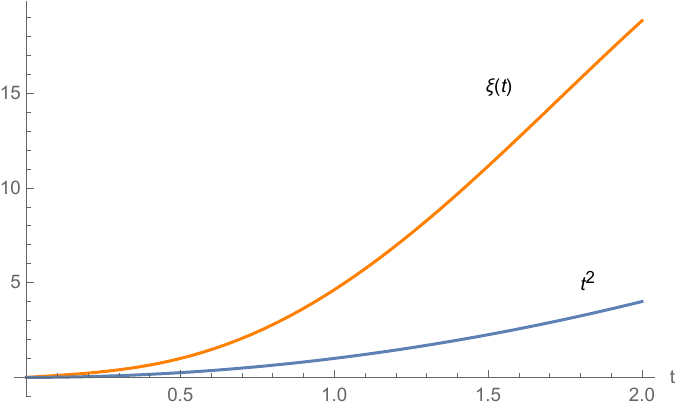}}
\caption{(a) Instability exponent near the equilibrium point $S(x_0,y_0,z_0)$ for the system Eq.(\ref{eq3.1}-\ref{eq3.3}) by Eq.(\ref{eq2.12}) and $\xi_{10}=10^{-10}$. (b) Variation of $\xi(t)$ and $t^2$ against $t$.}
\label{fig6}
\end{figure}

\renewcommand{\theequation}{4.\arabic{equation}}
\setcounter{equation}{0}

\section*{IV: STABILITY OF A SYSTEM WITH OUT FIXED POINT}
Let us consider a system of Sprott case A hidden attractor \cite{jafari2}, which is a special case Nose-Hoover \cite{William} system, pertinent to many natural natural phenomenon \cite{Posch}.

\begin{eqnarray}
\dot{x}&=y \label{eq4.1}\\
\dot{y}&=-x+yz \label{eq4.2}\\
\dot{z}&=1-y^2 \label{eq4.3}
\end{eqnarray}

\begin{figure}[ht]
\begin{center}
\hspace{-1.0cm}\includegraphics[width=0.55\textwidth, height=6.0cm]{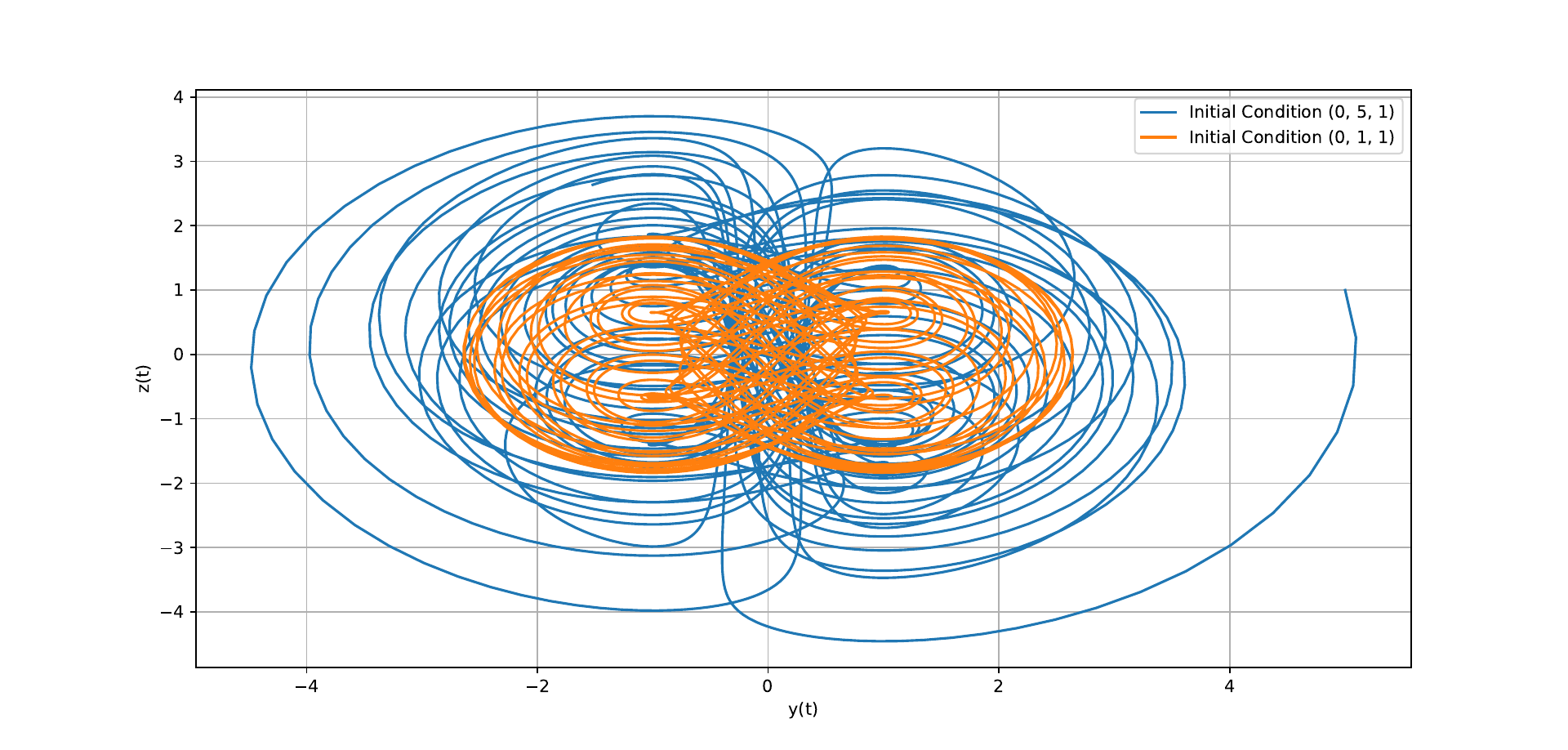}
\caption{Phase plot $(y-z)$ of the system Eq.(\ref{eq4.1}-\ref{eq4.3}) with two different initial conditions $(0,1,1)$ and $(0,5,1)$}
\label{fig7}
\end{center}
\end{figure}

Despite the inherent conservativity of the system, the absence of any apparent attractor is an expected outcome. However, our investigation uncovers a surprising and fascinating revelation – the presence of a coexisting chaotic sea alongside nested tori, each contingent on varying initial conditions Fig.(\ref{fig7}). This intriguing phenomenon strongly hints at the existence of a concealed attractor within the system.For the systematic and analytical prediction of the basin of attractor we employ KCC-theory as follows.

\begin{figure}[ht]
\begin{center}
\includegraphics[width=0.4\textwidth]{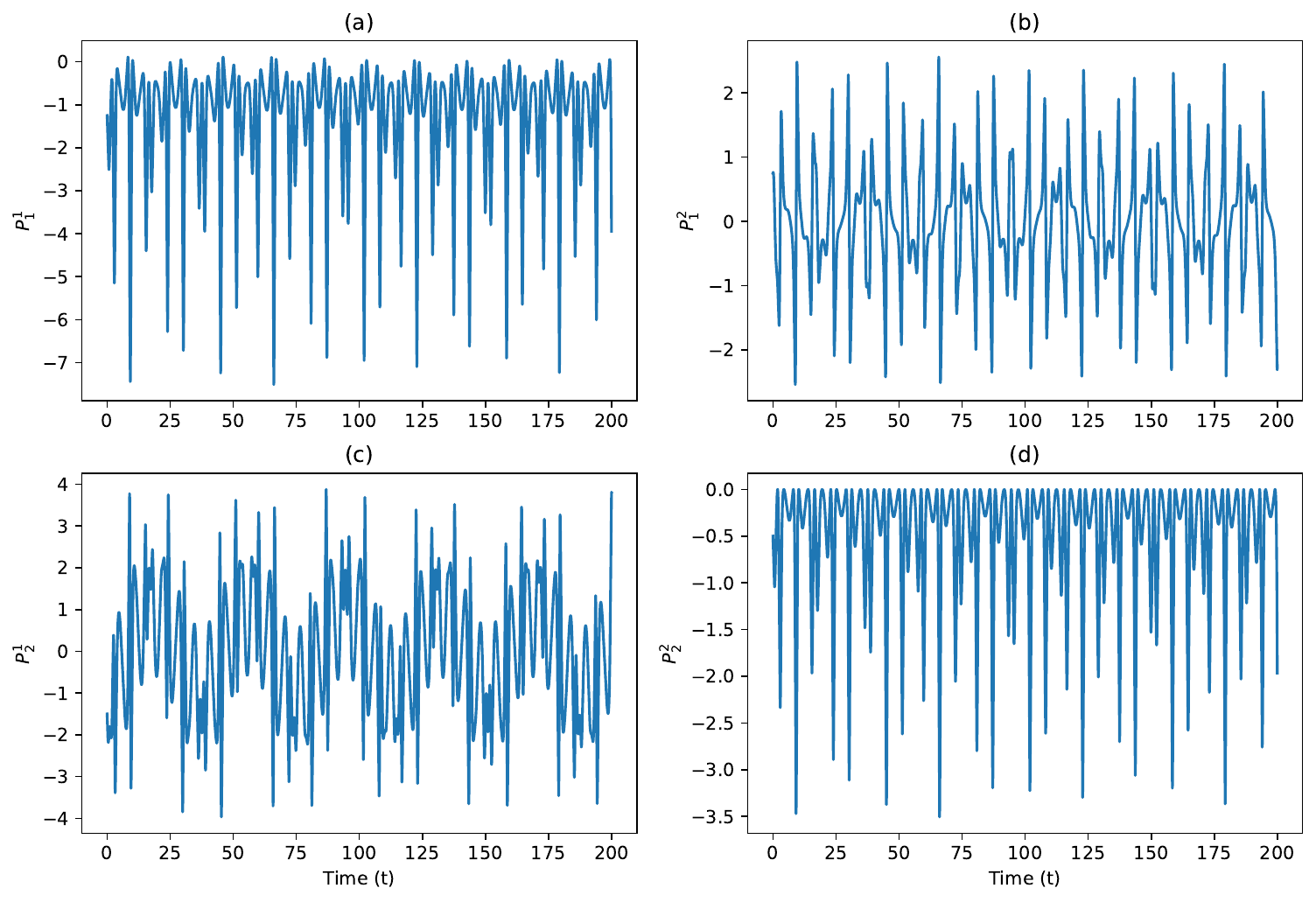}
\caption{Components of deviation curvature tensor $P_j^i$ Eq.(\ref{eq4.12}) of system Eq.(\ref{eq4.1}-\ref{eq4.3}) with initial condition (0,1,1) forming a nested tori. (a)$P_1^1$(t) vs time(t). (b)$P_2^1$(t) vs time(t). (c)$P_1^2$(t) vs time(t). (d)$P_2^2$(t) vs time(t)}
\label{fig8}
\end{center}
\end{figure}

By Eq.(\ref{eq3.2}) taking the derivative both sides and using $\dot{x}=y$ we arrive at
\begin{equation}
 \ddot{y}=-y+\dot{y}z+y\dot{z}  \label{eq4.4}
\end{equation}

also Eq.(\ref{eq3.3}) gives

\begin{equation}
 \ddot{z}=-2y\dot{y} \label{eq4.5}
\end{equation}

Now express the variable of the system by

\begin{equation}
\begin{split}
X^1&=y,~~~X^2=z,~~~X^3=x\\
Y^1&=\dot{y},~~~Y^2=\dot{z},~~~Y^3=\dot{x},
\label{eq4.6}
\end{split}
\end{equation}

one can symbolize Eq.(\ref{eq3.4} and \ref{eq3.5}) in comparison with Eq.(\ref{eq2.1}) as

\begin{eqnarray}
\frac{d^2X^1}{dt^2}+2G^1(X^1,X^2,Y^1,Y^2)=0 \label{eq4.7}\\
\frac{d^2X^2}{dt^2}+2G^2(X^1,X^2,Y^1,Y^2)=0 \label{eq4.8}
\end{eqnarray}

where the expression for $G^i$'s are specified as

\begin{eqnarray}
\begin{split}
G^1 &=\frac{1}{2}(X^1-Y^1X^2-X^1Y^2) \label{eq4.9}\\
G^2 &=X^1Y^1 \label{eq4.9}
\end{split}
\end{eqnarray}
Now the coefficients of nonlinear connection Eq.(\ref{eq2.3}) are calculated as

\begin{equation}
\begin{split}
\begin{cases}
N_{1}^1&=-\frac{1}{2}X^2,~~~N_{1}^2=-\frac{1}{2}X^1,\\
N_{2}^1&=X^1,~~~~N_{2}^2=0.
\label{eq4.10}
\end{cases}
\end{split}
\end{equation}

The components of of Berwald connection $G_{jl}^i=\frac{\partial N_{j}^i}{\partial Y^l}$ are worked out

\begin{equation}
\begin{split}
\begin{cases}
G_{11}^1&=G_{12}^1=G_{21}^1=G_{22}^1=0,\\
G_{11}^2&=G_{12}^2=G_{21}^2=G_{22}^2=0.
\label{eq4.11}
\end{cases}
\end{split}
\end{equation}

After successfully calculating the coefficients of nonlinear and Berwald connection, our analytical journey leads us to the computation of the coefficients of the deviation curvature tensor Eq.(\ref{eq2.9}). This pivotal step equips us with the tools necessary to make informed assessments about the Jacobi stability of the system.\\

Henceforth, the deviation curvature tensor, commonly recognized as the second KCC invariant, can be derived as follows:

\begin{equation}
\begin{split}
\begin{cases}
P_{1}^1&=\frac{1}{2}(Y^2)-1+\frac{1}{4}(X^2)^2-\frac{1}{2}(X^1)^2,\\
P_{2}^1&=\frac{1}{2}Y^1+\frac{1}{4}X^1X^2,\\
P_{1}^2&=-Y^1-\frac{1}{2}X^1X^2,\\
P_{2}^2&=-\frac{1}{2}(X^1)^2.
\label{eq4.12}
\end{cases}
\end{split}
\end{equation}

\begin{figure}[h]
\begin{center}
\includegraphics[width=0.4\textwidth]{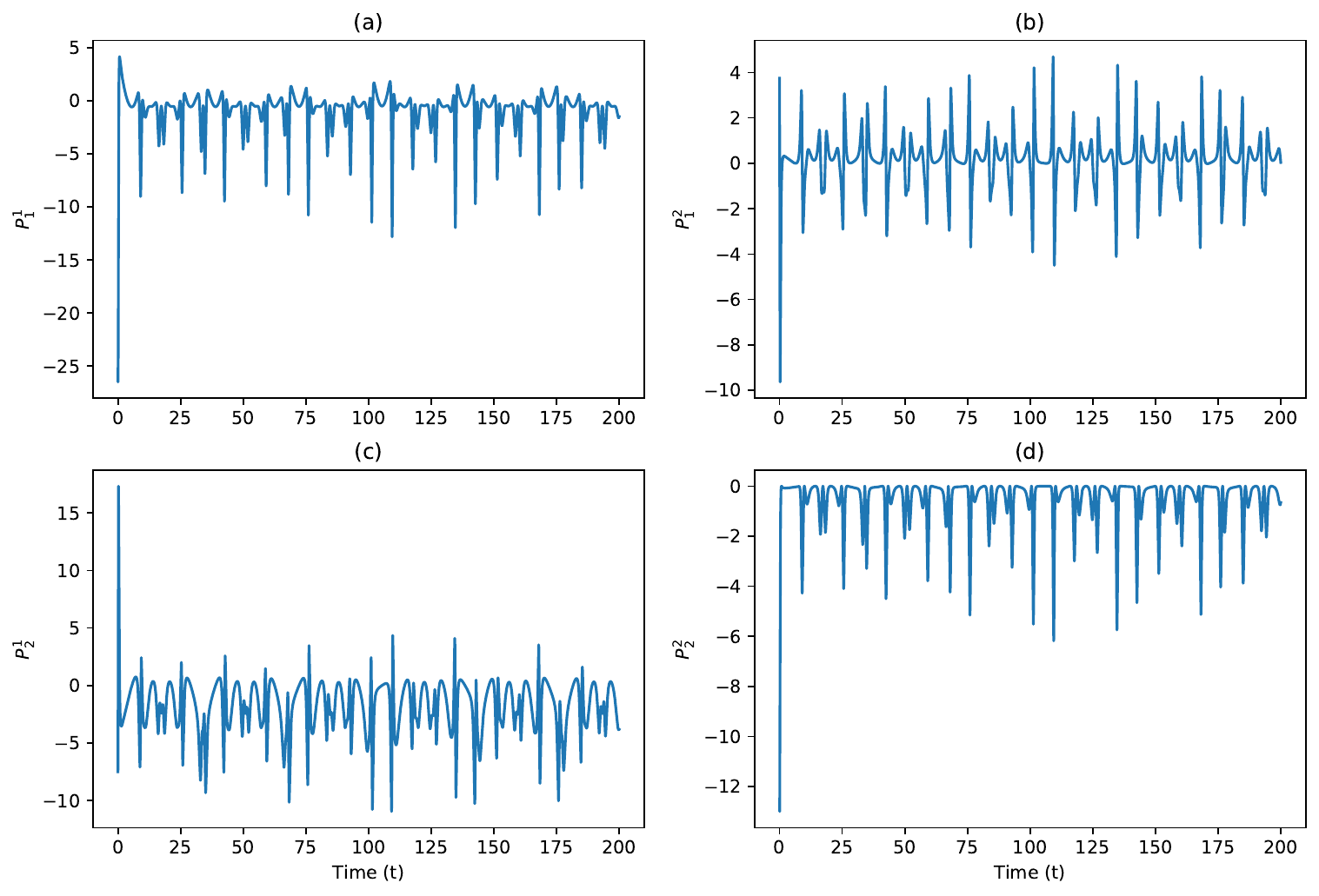}
\caption{Components of deviation curvature tensor $P_j^i$ Eq.(\ref{eq4.12}) of system Eq.(\ref{eq4.1}-\ref{eq4.3}) with initial condition (0,5,1) forming a chaotic attractor. (a)$P_1^1$(t) vs time(t). (b)$P_2^1$(t) vs time(t). (c)$P_1^2$(t) vs time(t). (d)$P_2^2$(t) vs time(t)}
\label{fig9}
\end{center}
\end{figure}

In Fig.(\ref{fig8} and \ref{fig9}), the temporal evolution of $P_i^j$ components is illustrated for identical initial conditions as previously employed in Fig.(\ref{fig7}). Specifically, the components of $P_i^j$ delineate the variation for the nested tori when initialized at $(0,1,1)$ and the chaotic sea for the initial condition (0,5,1) respectively.

Specifically the $P_{j}^i$ matrix is give by
$\begin{bmatrix}
P_{1}^1 & P_{1}^2\\
P_{2}^1 & P_{2}^2
\end{bmatrix}$
and its egenvalues can be acquired as

\begin{equation}
\lambda_{1,2}=\frac{1}{2}\big[P_{1}^1+P_{2}^2\pm \sqrt{(P_{1}^1-P_{2}^2)^2+4P_{1}^2P_{2}^1}\big],
\label{eq4.13}
\end{equation}
which satisfies the characteristics equation

\begin{equation}
\lambda^2-(P_{1}^1+P_{2}^2)\lambda+(P_{1}^1P_{2}^2-P_{2}^1P_{1}^2)=0.
\label{eq4.14}
\end{equation}
According to Routh-Hurwitz criteria, the system is Jacobi stable if

\begin{equation}
P_{1}^1+P_{2}^2<0 , ~~~~ P_{1}^1P_{2}^2-P_{2}^1P_{1}^2>0,
\label{eq4.15}
\end{equation}

otherwise it is Jacobi unstable.\\

From Eq.(\ref{eq4.13}), it becomes evident that the eigenvalues are solely determined by the coordinates $(X^1, X^2, Y^1, Y^2)$. Consequently, this insight provides us with a comprehensive visualization of the basin, particularly within the $(X^1-X^2)$ plane, where the eigenvalues can assume positive, negative, or complex values.

Crucially, it's noteworthy that the entire framework of the KCC theory hinges on the dynamics of nearby trajectories around initial points. This sets it apart from conventional linearization methods that necessitate prior knowledge of equilibrium points. Consequently, the KCC theory offers a more robust analytical approach for determining the basin of attraction, even in systems devoid of fixed points, as opposed to the arduous task of numerically evaluating Lyapunov exponents.

We take a significant step forward by proceeding to evaluate the deviation vector $\xi^i$, a decisive element that provides crucial insights into the onset and characterization of chaos within the system Eq.(\ref{eq4.1}-\ref{eq4.3}).One can write the differential equations of $\xi^i$ by Eq.(\ref{eq2.8})

\begin{align}
\frac{d^2\xi^1}{dt^2}&-X^2\frac{d\xi^1}{dt}-X^1\frac{d\xi^2}{dt}+(1-Y^2)\xi^1-Y^1\xi^2=0 \label{eq4.16}\\
\frac{d^2\xi^2}{dt^2}&+2X^1\frac{d\xi^1}{dt}+2Y^1\xi^1=0 \label{eq4.17}
\end{align}

Then by Eq.(\ref{eq4.6}) above equations can be written as
\begin{align}
\frac{d^2\xi^1}{dt^2}&-z\frac{d\xi^1}{dt}-x\frac{d\xi^2}{dt}+(1-\dot{z})\xi^1-\dot{x}\xi^2=0 \label{eq4.18}\\
\frac{d^2\xi^2}{dt^2}&+2x\frac{d\xi^1}{dt}+2\dot{x}\xi^1=0 \label{eq4.19}
\end{align}

\begin{figure}[ht]
\centering
\subfigure[]{\includegraphics[width=0.25\textwidth]{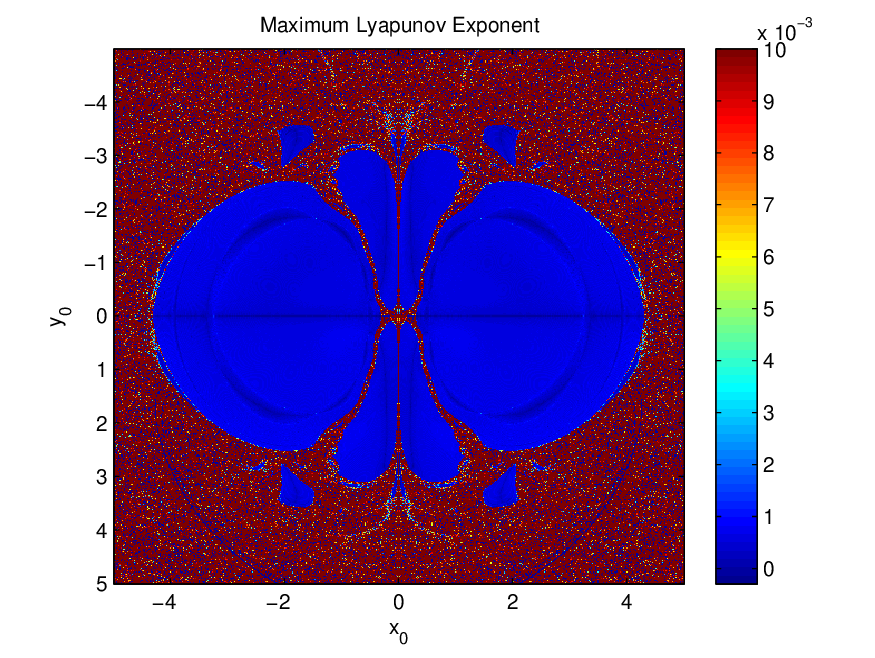}}\quad
\subfigure[]{\includegraphics[width=0.25\textwidth]{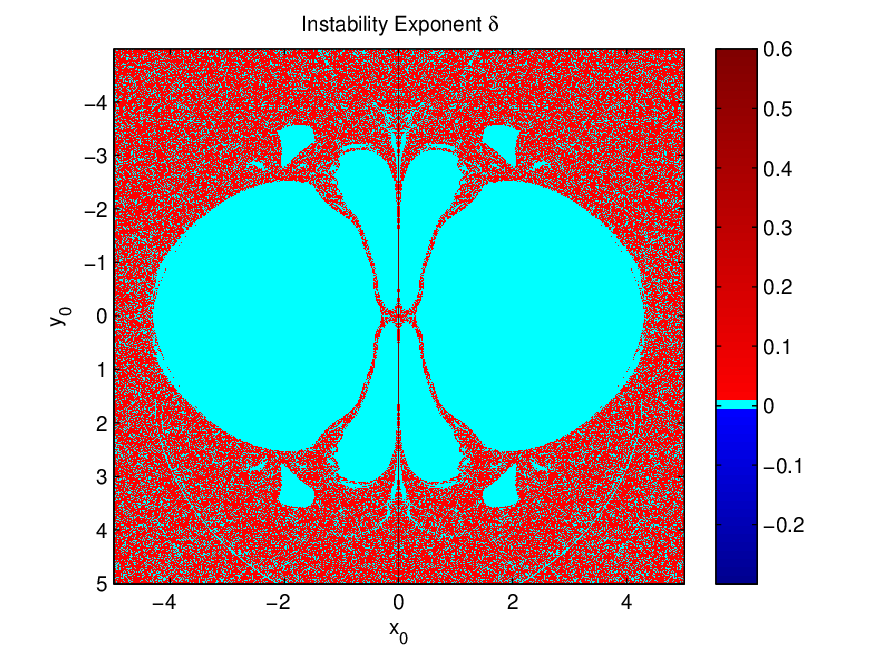}}
\caption{Basin of attraction of the system Eq.(\ref{eq4.1}-\ref{eq4.3}) (a)Basin ($x_0,y_0$) is numerically determined by the estimation of maximum Lyapunov exponent (b) Basin by calculating the instability exponent $\delta$ with help of deviation vector computed from Eq.(\ref{eq4.18}-\ref{eq4.19}). }
\label{fig10}
\end{figure}

Given that the system lacks equilibrium points, describing the behavior of deviation vectors through local stability analysis, as done in the previous case, is not applicable. However, by utilizing Eq. (\ref{eq4.18}) and (\ref{eq4.19}), we can numerically compute the instability exponent. This provides a means to characterize the basin of attraction Fig.(\ref{fig10}) for this type of hidden attractor.\\

\renewcommand\thesubfigure{(\alph{subfigure})}
\setcounter{subfigure}{0}

The presented graphical representation offers a compelling insight into the consistency between the numerically computed basin using the maximum Lyapunov exponent and the basin derived from the instability exponent obtained through the two analytically coupled differential equations. This observation holds true not only for the system with a stable equilibrium but also extends to the system without equilibrium points. The virtually perfect match in both scenarios attests to the robustness and versatility of our analytical approach. It underscores the effectiveness of deriving instability exponents from the coupled differential equations, offering a reliable and comprehensive understanding of the basin of attraction, even in systems lacking equilibrium points.

\renewcommand{\theequation}{5.\arabic{equation}}
\setcounter{equation}{0}

\section*{V: CONCLUSIONS}

In conclusion, our exploration has provided profound insights into the intricate dynamics of two distinct autonomous 3-D systems. The first system, characterized by a singular stable fixed point, exhibited unexpected chaotic behavior, challenging conventional expectations from linear stability analysis. Utilizing the Kosambi-Cartan-Chern (KCC) theory, we successfully captured the basin of attraction for this system through the analytical depiction of deviation vectors and curvature variation, unveiling hidden chaotic oscillations.

Equally intriguing was the examination of the second system, which lacked equilibrium points altogether. By numerically computing the instability exponent using equations (\ref{eq4.18}) and (\ref{eq4.19}), we effectively characterized the basin of attraction for this unique hidden attractor. This analytical approach provides a valuable alternative to the conventional numerical evaluation of Lyapunov exponents, demonstrating the robustness and versatility of the KCC theory.

The comprehensive visualization of basins, as demonstrated in Fig.(\ref{fig4}) and Fig.(\ref{fig5}), not only substantiates the efficacy of the instability exponent in predicting chaotic zones but also underscores the superiority of the KCC theory in offering analytical insights into the behavior of complex dynamical systems. This work contributes significantly to the understanding of hidden attractors, showcasing the KCC theory as a powerful tool for capturing nuanced dynamics, even in scenarios where equilibrium points are absent. Our findings open avenues for further exploration in the realm of nonlinear dynamics and provide a solid foundation for future research endeavors aimed at unravelling the complexities of chaotic systems like network dynamics, collective dynamics etc.

\section*{ACKNOWLEDGEMENT}
S.Roy expresses gratitude to Dr.Tanmoy Paul for his profound insights and valuable discussions during the course of this work.

\end{document}